\documentclass[12pt]{iopart}

\usepackage{graphicx}
\usepackage{subfigure}
\usepackage{url}
\usepackage{color}

\begin{document}

\title[RSA of aligned rectangles with two discrete orientations]{Random sequential adsorption of aligned rectangles with two discrete orientations: Finite-size scaling effects}

\author{Luca Petrone}
 \address{via B. Cesana 6, 20132 Milano, Italy}
\author{Nikolai Lebovka}
\address{Laboratory of Physical Chemistry of Disperse Minerals, F. D. Ovcharenko Institute of Biocolloidal Chemistry, NAS of Ukraine, Kyiv 03142, Ukraine}
\author{Micha\l{} Cie\'sla}%
\address{Institute of Theoretical Physics, Jagiellonian University, 30-348 Krak\'ow, \L{}ojasiewicza 11, Poland}%

\date{\today}

\begin{abstract}
We study saturated packings produced according to random sequential adsorption (RSA) protocol built of identical rectangles deposited on a flat, continuous plane. An aspect ratio of rectangles is defined as the length-to-width ratio, $f=l/w$. The rectangles have a fixed unit area (i.e., $l \times w=1$), and therefore, their shape is defined by the value of $f$ ($l=\sqrt{f}$ and $w=1/\sqrt{f}$). The rectangles are allowed to align either vertically or horizontally with equal probability. The particles are deposited on a flat square substrate of side length $L$ (measured in units of particle length, $L \in [20, 1000]$) and periodic boundary conditions are applied along both directions. The finite-size scaling effects are characterized by a scaled anisotropy defined as $\alpha = l/L = \sqrt{f}/L$. We showed that the properties of such packings strongly depend on the value of aspect ratio $f$ and the most significant scaling effects are observed for relatively long rectangles when $l\ge L/2$ (i.e. $\alpha \ge 0.5$). It is especially visible for the mean packing fraction as a function of the scaled anisotropy $\alpha$. The kinetics of packing growth for low to moderate rectangle anisotropy is to be governed by $\ln t/t$ law, where $t$ is proportional to the number of RSA iterations, which is  the same as in the case of RSA of parallel squares. We also analyzed global orientational ordering in such packings and properties of domains consisting of a set of neighboring rectangles of the same orientation, and the probability that such domain forms a percolation. 
\end{abstract}

\noindent{\it Keywords\/}: random sequential adsorption; packing of aligned rectangles; kinetics of packing growth; percolations
\maketitle

%
%
%
\section{Introduction}
Random sequential adsorption (RSA) is one of the protocols used for generating random packings. The next particle with a random position and orientation is added to the packing only if it does not intersect with any of the previously deposited objects. Otherwise, it is removed and abandoned. The added particles do not change positions and orientations. The packing generation finishes when there is no place large enough to add the next object, and such packing is called saturated. Although the first application of the RSA protocol dates from 1939 when Flory studied the attachment of blocking groups on a linear polymer chain \cite{Flory1939}, RSA owes its name and popularity to Feder, who noticed that such packings resemble monolayers formed in irreversible adsorption processes \cite{Feder1980}. Since then RSA become an important tool in the modeling of adsorption monolayers \cite{Kubala2022, Talbot2000}, but, because of its simplicity, it attracted a lot of attention from theoretical studies devoted to properties of random media \cite{Evans1993, Torquato2010}. Besides the protocol seeming to be very simple, there are not many analytical results, especially for continuous packings. For example, a mean saturated packing fraction, which is the most important property of packing, was determined analytically only in the case of RSA of segments on a one-dimensional line \cite{Renyi1958} and equals $0.7475979202...$. The same parameter for two-dimensional RSA packing built of parallel squares was determined numerically only and equals $0.562009(4)$ \cite{Brosilow1991}. The situation changes when rectangles' orientations are arbitrary. For unoriented squares, the packing fraction equals $0.527640(18)$ \cite{Kasperek2018, Zhang2018}, and for unoriented rectangles, it reaches the maximum at $0.549632(17)$ for a side length ratio of $1.492(22)$ and then decreases monotonically with the growth of side length ratio \cite{Vigil1989, Kasperek2018}. It is worth noting, that the highest RSA packing fraction for this kind of system was recently reported for horizontally or vertically aligned rectangles with random side length ratio but constant surface area \cite{Petrone2021, Petrone2022}. There the packing fractions were up to $0.7$, depending on the probability distribution used for choosing rectangles' anisotropy. 

Another property is the kinetics of packing growth. Generally, it is described by the power law
\begin{equation}
\label{eq:fl}
    \theta - \theta(t) \sim t^{-\frac{1}{d}},
\end{equation}
where $\theta(t)$ is the packing fraction after $t$ trials of adding a new particle to the packing. Here, $\theta \equiv \theta(t\to\infty)$ is the saturated packing fraction \cite{Feder1980, Pomeau1980, Swendsen1981}. The parameter $d$ is related to the number of particle degrees of freedom \cite{Hinrichsen1986, Ciesla2013}. Among others, such behavior was also reported for RSA of freely oriented rectangles of fixed aspect ratio \cite{Vigil1989} and for horizontally or vertically aligned ones with random aspect ratio but fixed area \cite{Petrone2021, Petrone2022}. 
On the other hand, for RSA of parallel squares, 
the kinetics of packing fraction takes the form \cite{Brosilow1991}
\begin{equation}
\label{eq:log}
    \theta - \theta(t) \sim \frac{\ln (t)}{t},
\end{equation}
It is worth noting, that for lattice RSA the packing growth kinetics follows another, exponential law
\begin{equation}
\label{eq:exp}
    \theta - \theta(t) \sim \exp\left(-\frac{t}{\sigma}\right),
\end{equation}
where $\sigma$ depends on lattice details \cite{Privman1991}.

The third aspect of random arrangements of anisotropic shapes like rectangles is their tendency to form orientationally oriented clusters. Typically in RSA packings, the range of orientational ordering is very short \cite{Kasperek2018, Ricci1992, Abritta2022, Sherwood1997}, however, for very elongated molecules it can be significant \cite{Ciesla2013}, and orientationally oriented domains can appear \cite{Barbasz2013}. It appears that especially in lattice packings, where orientations are discrete, one can observe long-range connectivity, which could lead to percolation transition, typically observed and studied for RSA on lattices \cite{Kondrat2001, Cadilhe2007, Lebovka2011, Tarasevich2015, Centres2015, Budinski2016, Tarasevich2018}.

To address all these phenomenons, we focus on the numerical study of RSA of rectangles placed on a continuous and flat plane that have two possible orientations: horizontal and vertical ones, but, in contrast to recent studies~\cite{Petrone2021, Petrone2022}, fixed aspect ratio. Thus, the studied packings could allow us to effectively address all the abovementioned properties at the border between lattice and continuous packings built of oriented and unoriented elongated particles. Additionally, we also studied all these properties in the limit of very long elongations that makes the length of the longer rectangle side comparable with packing size.
\section{Algorithm}
\label{sec:algorithm}
\begin{figure}
    \centering
    \includegraphics[width=0.7\columnwidth]{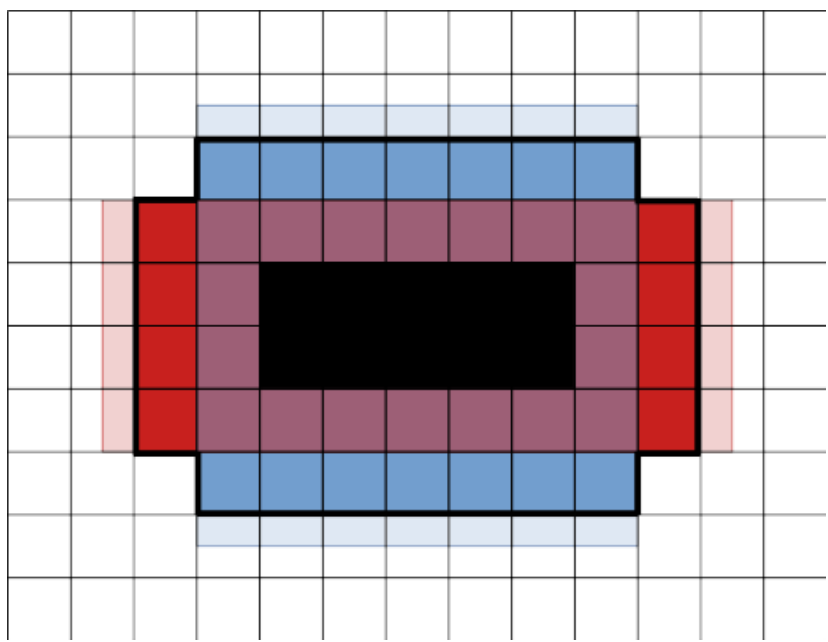}
    \caption{Area blocked by the deposited black rectangle. The red region is blocked for placing the center of a horizontally aligned rectangle and the blue one is for the center of a vertically aligned rectangle. The whole packing is divided into smaller squares. A square is available for further deposition of a rectangle of a given orientation if it is not fully covered by the corresponding blocked area. The available squares are outside the bold black line.}
    \label{fig:excludedzones}
\end{figure}

The algorithm for generating strictly saturated packing built of horizontally and vertically oriented rectangles bases on the concept introduced by Wang~\cite{Wang1994} of tracing regions large enough to hold a next shape. In order to implement it, we divide the whole packing area into squares, and identified, which of them are available for further deposition. Each deposited rectangle excludes the areas shown in Fig.~\ref{fig:excludedzones}
Because there are two possible orientations we have two separate lists for these squares. One for the possible placing of horizontally aligned rectangles and the other for vertically aligned ones. If an area is completely covered by a union of the exclusion zones of deposited rectangles, it is not available for further depositions and is removed from the corresponding list. Thus, the available squares approximate regions where centers of subsequent rectangles can be placed. After a certain number of trials of adding a new particle to the packing the available squares are divided - each one into four smaller squares of a two times smaller side length, and the new squares are reexamined if they are still available. It allows for improving the approximation of available regions.

In detail, the crucial part of the algorithm that determines if the square is available or not consists of a sequential evaluation of the points described below.
\begin{figure}
    \centering
    \includegraphics[width=0.7\columnwidth]{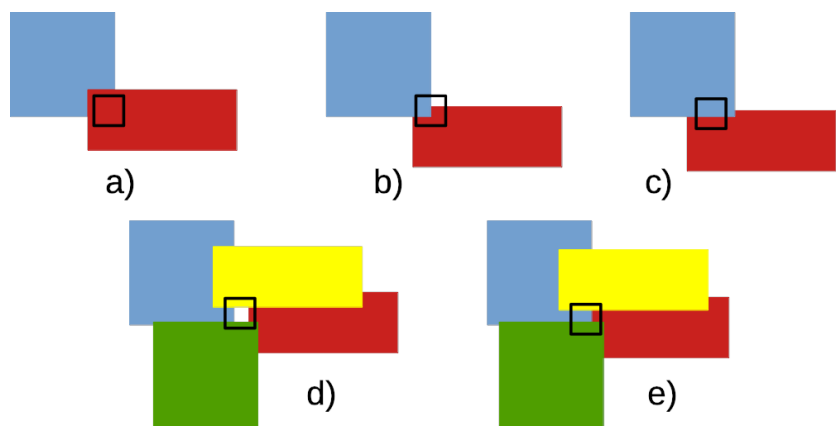}
    \caption{Various relative positions of analyzed square (with black sides) and excluded regions of neighboring rectangles. In cases a), c), and e), the square is fully covered and can be removed from the list of available squares. In cases b) and d), it contains the area where it is possible to place the center of the next rectangle.}
    \label{fig:boxes}
\end{figure}
\begin{enumerate}
\item Find neighboring rectangles to the current square. A rectangle is neighboring if its excluded region overlaps with the square.
\item If all of the four vertices of the square are inside an excluded region of a single rectangle, then the square is fully covered (Fig.~\ref{fig:boxes}a), and is removed from the corresponding list.
\item Otherwise, if any one of the square vertices is not covered by any neighboring rectangle excluded region, the square is not fully covered (Fig.~\ref{fig:boxes}b), and it stays on the list of available squares.
\item Otherwise, if all square edges are completely covered by two or three neighboring rectangles excluded regions, the square is fully covered (Fig.~\ref{fig:boxes}c), and it is removed from the corresponding list. To check if the edge is covered we used Klee's algorithm \cite{Klee1977}. 
\item Otherwise, we know that the square may be covered by at least four excluded regions. Even if all its edges are covered, the square can still contain uncovered area (Fig.~\ref{fig:boxes}d). To check it, the coordinates of the intersection points of the excluded regions are calculated. If each intersection point of any pair of the excluded regions is covered by any other region, the square is fully covered (Fig.~\ref{fig:boxes}e, and is removed from the corresponding list. Otherwise, it is not fully covered and it stays on the list.
\end{enumerate}

\subsection{Numerical details}

Rectangles with length $l$, width $w$, and unit area (i.e. $l \times w=1$) are deposited on a flat square substrate of side length $L \in [30, 1000]$. Periodic boundary conditions are used to decrease finite size effects \cite{Ciesla2018}. For a given packing size, typically $100$ independent packings were generated. The position of the deposited figure is drawn according to the uniform probability distribution. The shape of a rectangle is determined by the aspect ratio defined as the length-to-width ratio, $f = l/w$. Therefore, we have $l=\sqrt{f}$ and $w=1/\sqrt{f}$. Horizontal and vertical orientations are equally probable. For a given $f$ and $L$, the scaled anisotropy was defined as
\begin{equation}
\label{eq:alpha}
    \alpha = \frac{l}{L} = \frac{\sqrt{f}}{L}.
\end{equation}

To speed up the overlap detection of rectangles we implemented a neighbor grid that covers the whole packing. The grid is built of square cells. Each cell contains a list of already deposited rectangles with the center inside it. The cell size is $2l \times 2l$, thus, the rectangle can only intersect with another shape from the same cell or from one of its eight nearest neighbors. The newly deposited rectangle finds all the cells it covers and then checks overlap with all the rectangles stored in these cells' lists. 

After the initial filling of packing with rectangles, when the number of consecutive unsuccessful trials of placing a new figure reaches a threshold value (we used $10^3$), we initiate the square lists described in the previous section, and further attempts are restricted only to available squares.

It should be noted that the aspect ratio $f$ should not exceed $L^2$ (i.e. $\alpha \le 1$), to avoid self-overlapping due to periodic boundary conditions.
\section{Results}
\begin{figure}
    \centering
    \includegraphics[width=0.3\columnwidth]{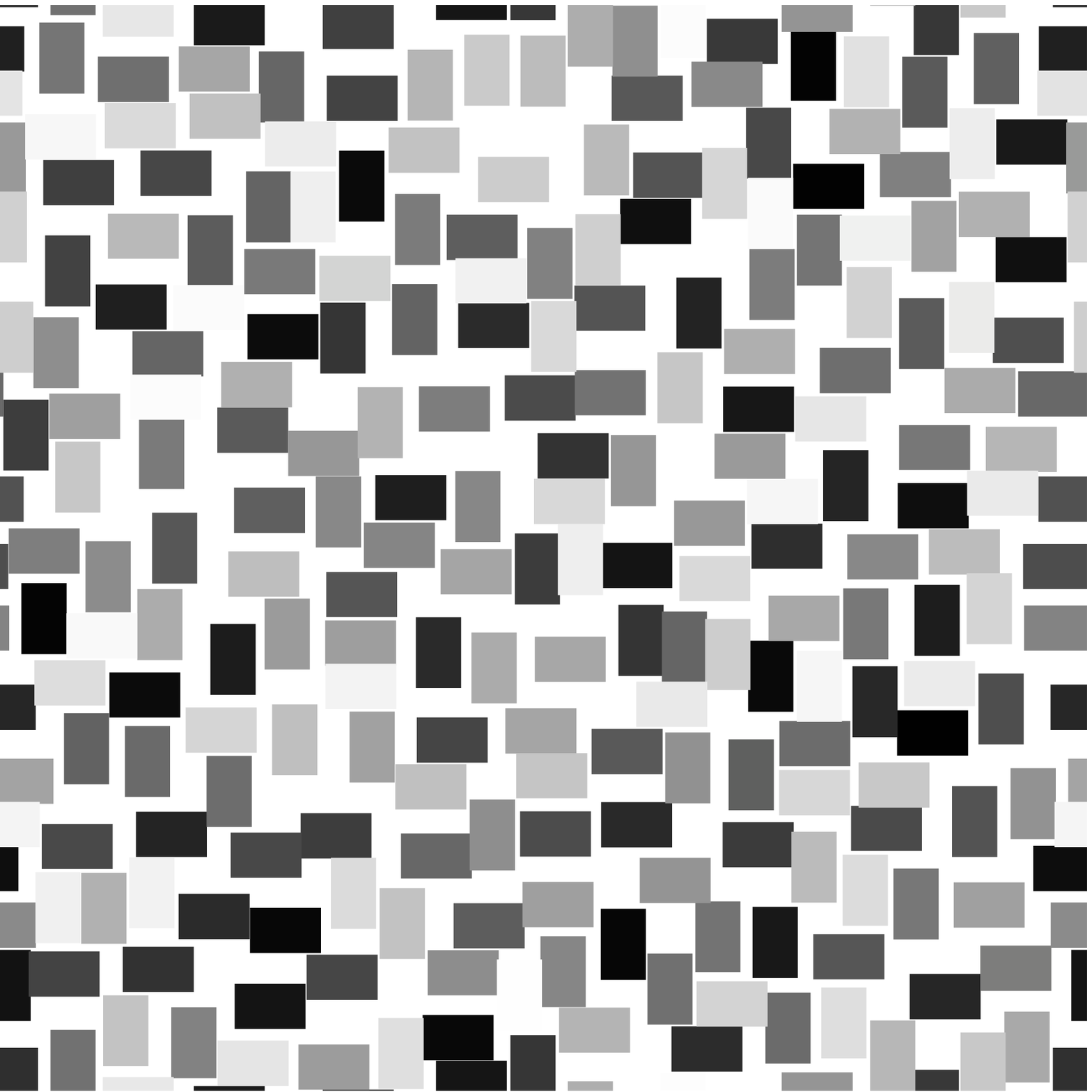}
    \includegraphics[width=0.3\columnwidth]{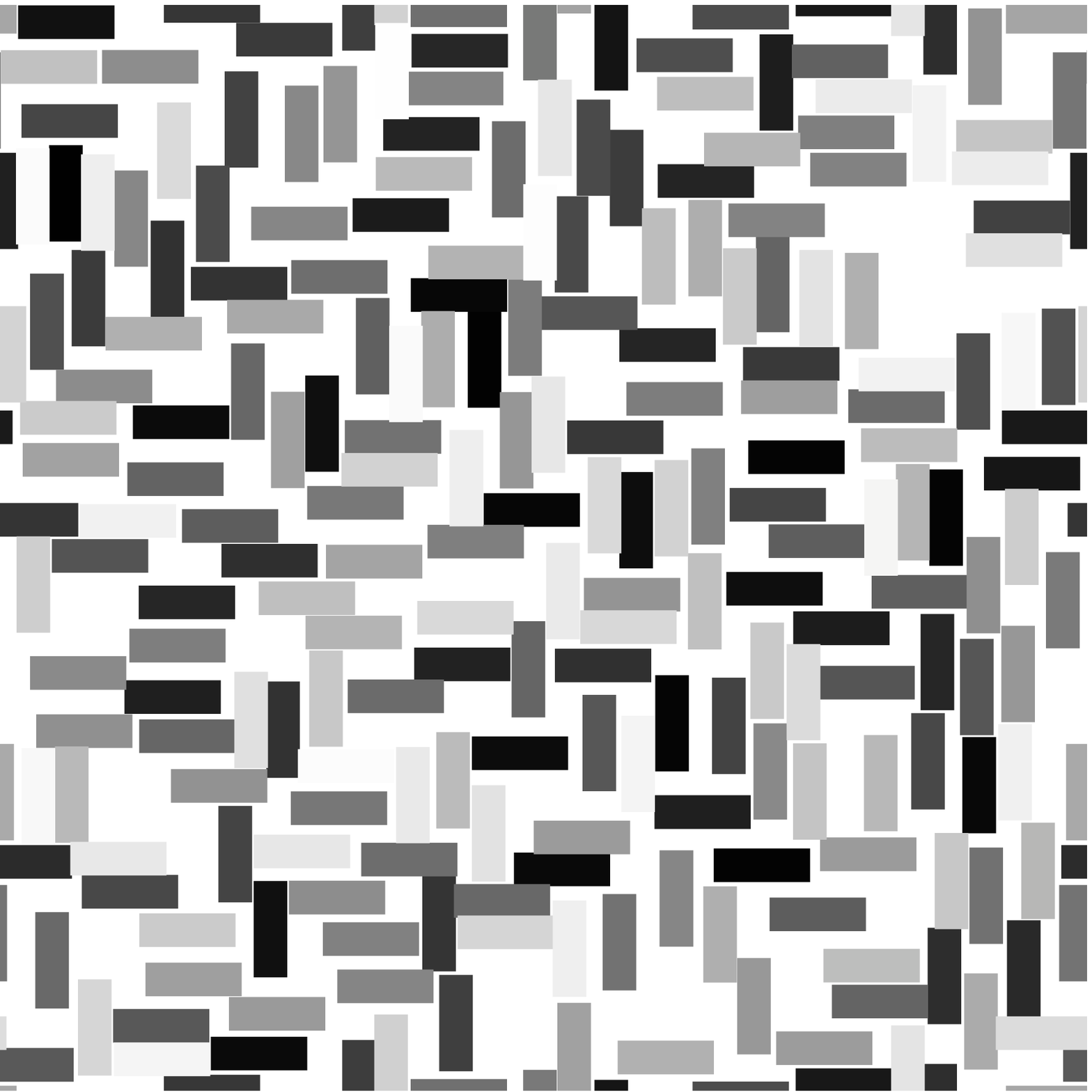}
    \includegraphics[width=0.3\columnwidth]{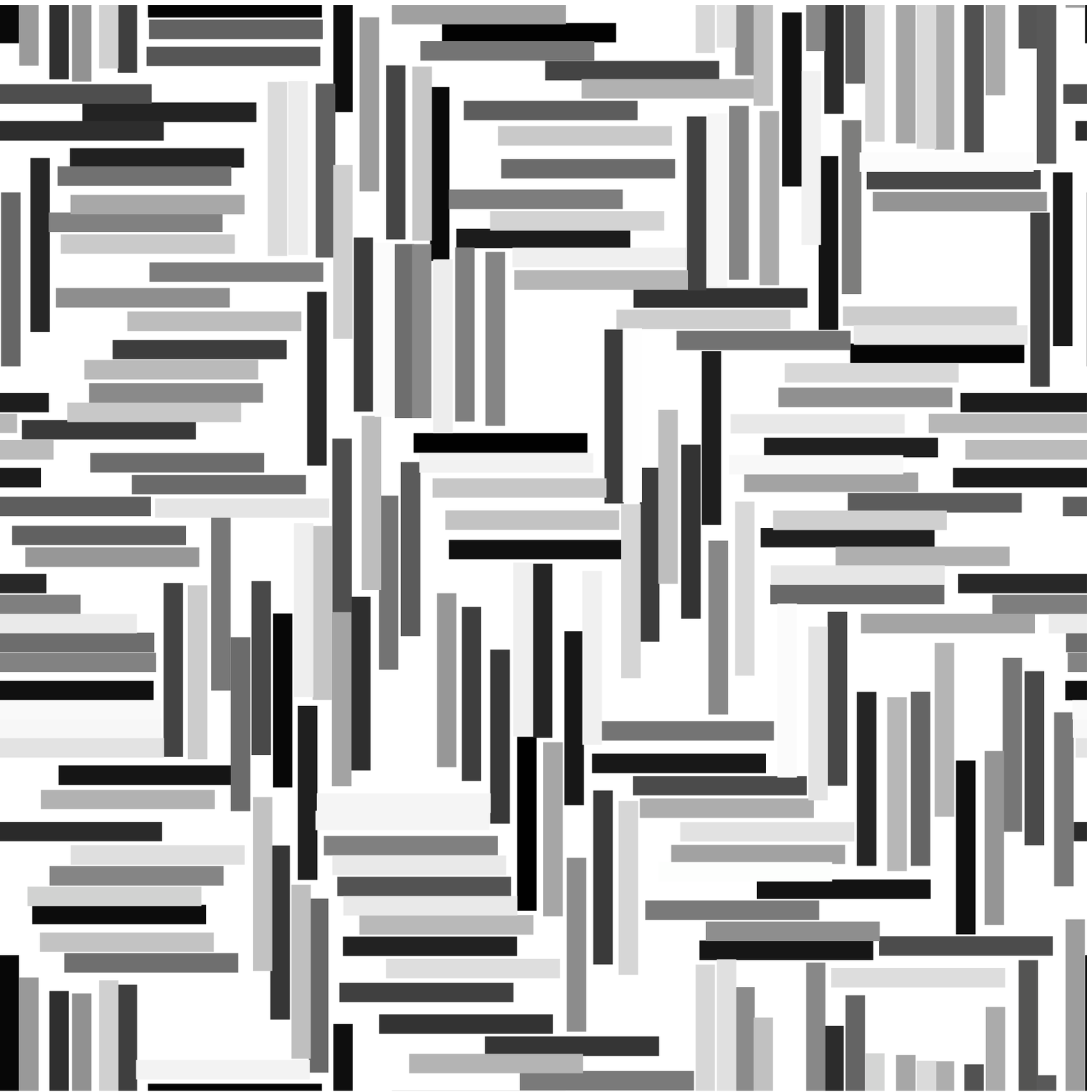}
    \caption{Example saturated packings of $L=20$ build of rectangles of aspect ratio $f=1.6, 3.0$, and $10.0$ for left, middle and right panel, respectively. Different colors of rectangles are to better distinguish them visually.}
    \label{fig:examples}
\end{figure}

Fragments of example packings are shown in Fig.~\ref{fig:examples}.
The properties of obtained packing are presented and discussed in the following sections.
\subsection{Mean saturated packing fraction}
\begin{figure}[ht]
 \centering
 \includegraphics[width=0.8\columnwidth]{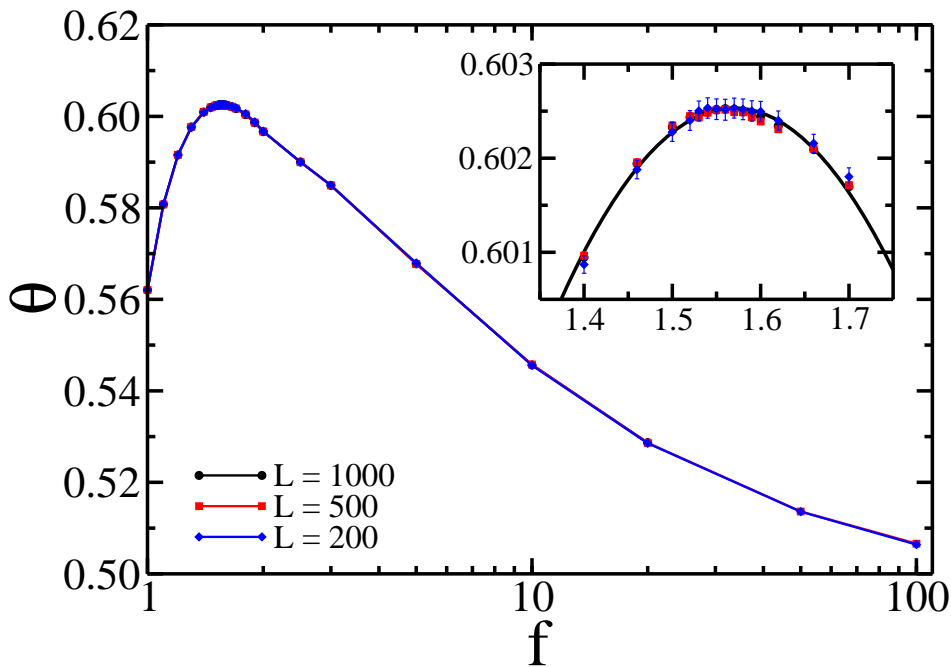}
 \caption{The mean saturated packing fraction as a function of the aspect ratio $f$ for several values of packing size $L$. Points correspond to measured values with error bars denoting their standard deviation. Lines are to guide the eye. The inset shows the magnification near the maximum. The solid line is the least squares fit to the data for $L=1000$: $\theta = 0.60254 - [0.23(f-1.57)]^2$.}
 \label{fig:theta_f}
\end{figure}

The mean saturated packing fraction $\theta$ for small and moderate aspect ratio $f$ and sufficiently large $L$ ($=1000$) is shown in Fig.~\ref{fig:theta_f}.
First, the mean saturated packing fraction of squares aligned in parallel is $0.562019 \pm 0.000021$ and it agrees with the earlier report \cite{Brosilow1991, Wang2000}. The packing fraction grows with the rectangle anisotropy up to the value of $\theta_{max} = 0.60252694 \pm 0.000019$ reached for $f=1.56$. These values were calculated by averaging saturated packing fractions of $100$ independently generated packings of size $L=1000$. The position of the maximum and its value does not depend on the size of the packing, which means that finite-size effects do not affect presented results. It is quite reasonable for the selected value of $f$ ($f\le 100$ and $L \ge 200$), i.e. very small values of scaled anisotropy $\alpha$ ($\le 0.05$). 
The presence of this maximum at low to moderate sides-ratio is the effect of competition between two factors. The first is the tendency to place figures in parallel at the late stages of packing generation which promotes parallel alignment and favors denser configuration. On the other hand, anisotropy spoils the packing density at the beginning of packing generation because the mean area blocked by a single shape grows with its sides-ratio. Therefore there is an optimal value of anisotropy for which packing fraction is maximized \cite{Vigil1989}. This effect
was observed for a number of packings built of anisotropies objects \cite{Viot1992, Sherwood1997, Haiduk2018, Ciesla2016}. Interestingly, for RSA of unoriented rectangles, the maximum packing fraction was observed for sides-ratio of $1.492 \pm 0.022$ \cite{Kasperek2018}, which is a close, but slightly smaller value than the one observed here. It is probably due to only two possible orientations of a rectangle in the packing, thus, the larger anisotropy does not harm the denser configuration as much as for unoriented shapes.

\begin{figure}[ht]
 \centering
 \includegraphics[width=0.8\columnwidth]{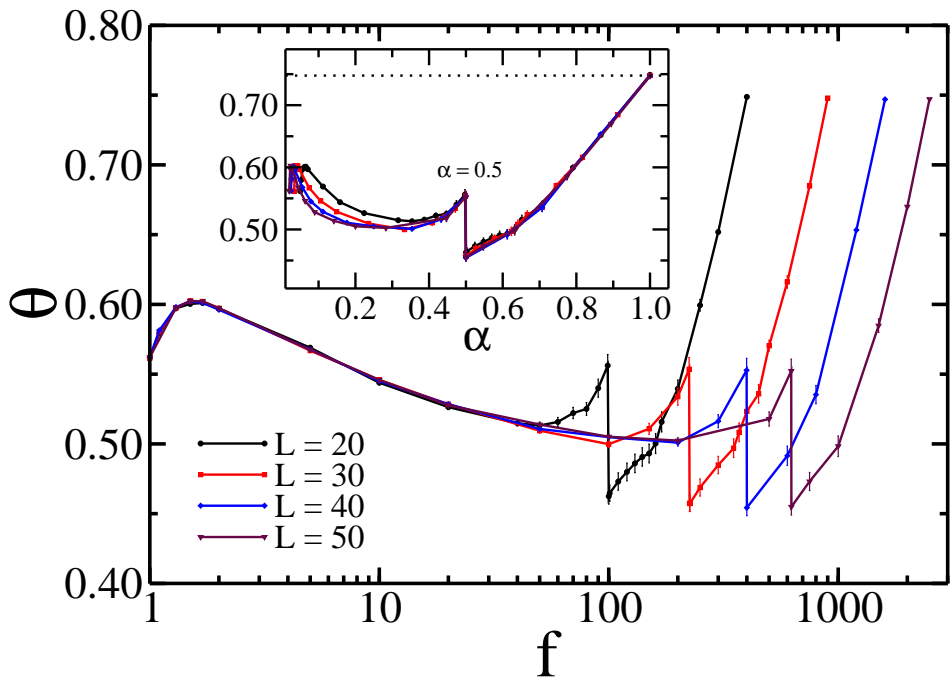}
 \caption{The mean saturated packing fraction as a function of the aspect ratio $f$ for smaller packings and the whole possible range of anisotropies. Dots are data measured from simulations and lines are to guide the eye. Inset shows the same data but rescaled - the control parameter on the horizontal axis is the ratio of the longer rectangle side length and packing side length. The dotted line corresponds to the saturated packing fraction in one dimension $\theta=0.7476...$}
 \label{fig:theta_f_long}
\end{figure}

For larger rectangle side-ratio, the packing fraction decreases, similarly as for RSA of unoriented anisotropic shapes \cite{Vigil1989, Kasperek2018, Haiduk2018, Viot1992}. However, for larger values of the scaled anisotropy $\alpha$ (Eq.~(\ref{eq:alpha})), when the particle length $l$ is comparable with the system size $L$, the orientation of the first-placed rectangle determines orientations of all the rest rectangles as there is no possibility of placing perpendicular figures without intersection with the first one. Thus, the process is effectively reduced to one-dimensional RSA, mentioned in the introduction section, for which the packing fraction is $0.7475979202...$ \cite{Renyi1958}. Therefore for finite packings, as in this study, we expect that the saturated packing fraction should begin to grow for some large value of $f$. To study this phenomenon we used relatively small packings because the computational complexity of the simulation algorithm described in Sec.~\ref{sec:algorithm} grows rapidly with rectangle anisotropy. The obtained results are shown in Fig.~\ref{fig:theta_f_long}.
Although the packing size is small, the dependence of packing fraction on rectangle anisotropy for small and moderate $f$ is the same as for larger packings. However here, the packing fraction starts to grow when the longer rectangle side length exceeds one-third of the packing side length. The local minimum packing fraction there is around $\theta=0.5$ and depends on the packing size - for smaller packings, this minimum is shallower. The growth is due to better filling the space of two rectangles laying one after the other or one above the other, and it continues up to the scaled anisotropy $\alpha = 0.5$ and $\theta \approx 0.55$. When this limiting value is exceeded, the sudden drop is observed to $\theta \approx 0.45$ because now two rectangles cannot be placed one after the other or one above the other, because they will overlap due to periodic boundary conditions. If there was only one possible orientation, the packing fraction would drop to half of the saturated packing fraction for one-dimensional RSA, but the possibility of placing figures perpendicular raises the packing density. It is worth noting, that the observed densities around $\alpha \approx 0.5$ are weakly or not dependent on packing side length $L$. For larger anisotropies, we observe monotonic growth of packing fraction, and for scaled anisotropy $\alpha \to 1$, as expected, we restore the saturated density of one-dimensional RSA packings.
\subsection{Kinetics of packing growth}
\begin{figure}[ht]
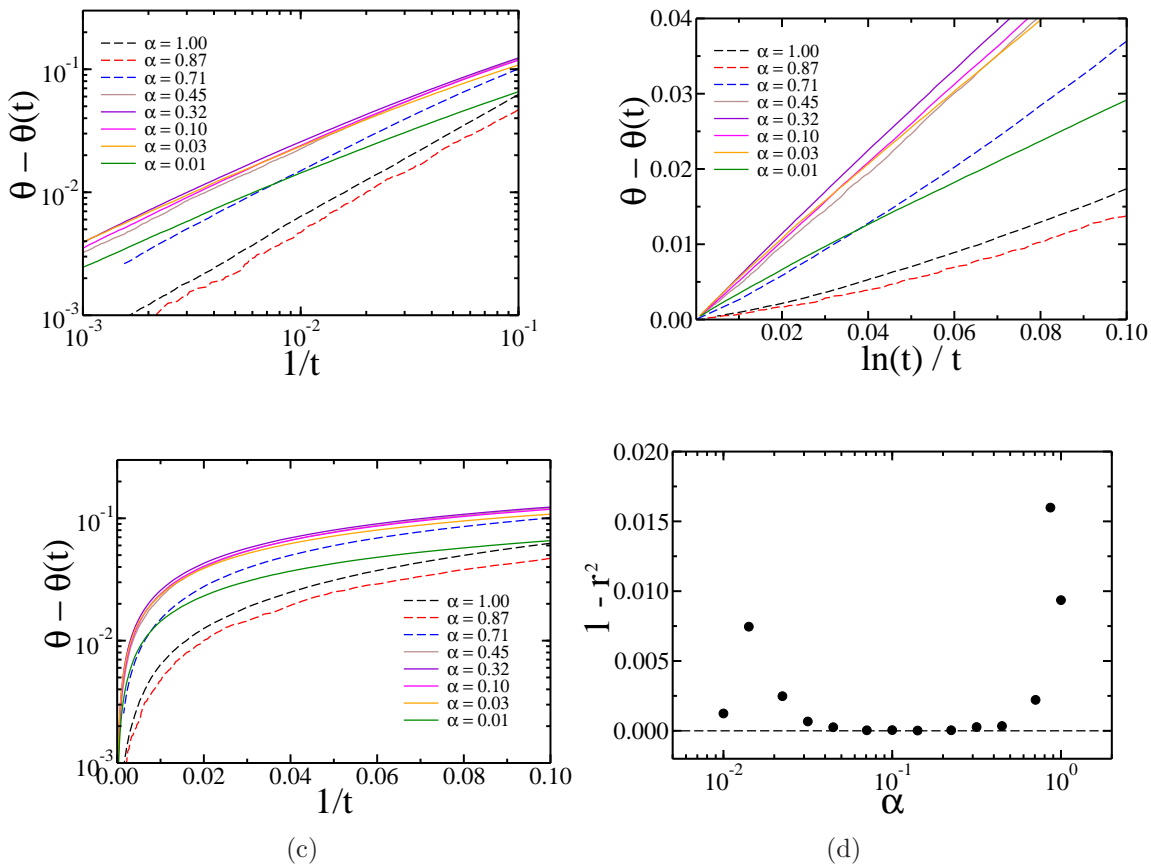

 \centering
 \subfigure[]{\includegraphics[width=0.45\columnwidth]{kinetics_100_pow.eps}}
 \hspace{0.05\columnwidth}
 \subfigure[]{\includegraphics[width=0.45\columnwidth]{kinetics_100_log.eps}}
 \subfigure[]{\includegraphics[width=0.45\columnwidth]{kinetics_100_exp.eps}}
 \subfigure[]{\includegraphics[width=0.45\columnwidth]{fit_log_r2.eps}}
 \caption{The kinetics of packing growth for $L=100$ and various scaled anisotropy $\alpha$. All plots show the same data, but presented in a different way to show differences between theoretical laws (\ref{eq:fl}) in (a), (\ref{eq:log}) in (b), and (\ref{eq:exp}) in (c). If the law is fulfilled the data should form a straight line for large enough $t$ (small $1/t$). Data for packings built of rectangles of the highest anisotropies were drawn using dashed lines for convenience. Panel (d) shows the dependence of linear fit accuracy for the data shown in panel (b). The dashed line corresponds to ideal linear dependence. The data was taken for $L=100$ and $100$ independently generated saturated packings. }
 \label{fig:kinetics_comparison}
\end{figure}

Kinetics of packing growth was studied in order to determine whether it obeys the power (\ref{eq:fl}), extended logarithmic (\ref{eq:log}), or exponential (\ref{eq:exp}) law. The first one is known to be valid for most cases of RSA \cite{Shelke2007} while the second was confirmed for RSA of parallel squares \cite{Brosilow1991}. The third one, as mentioned in the Introduction, is observed for RSA on lattices, where positions of packed objects are restricted to a finite (or countable in the limit of infinite packing size) number of places. The comparison of these theoretical laws for the exemplary data taken from $100$ independent packings at the limit of large $t$ for $L=100$ and $f \in [1, 10000]$, which corresponds to $\alpha \in [0.01 - 1]$ is shown in Fig.~\ref{fig:kinetics_comparison}.
The exponential law (\ref{eq:exp}) does not fit well for any anisotropy, as the dependence of $\theta - \theta(t)$ does not form a straight line -- see Fig.~\ref{fig:kinetics_comparison} (c). This is expected since such kinetic dependence has been observed so far only in lattice packings. On the other hand, for large enough $t$ all the data fit well to the power law (\ref{eq:fl}) -- see Fig.~\ref{fig:kinetics_comparison} (a). The fit correlation coefficient exceeds $0.99$. This is because, for large but finite $t$, which is the case in computer simulations, the $1/t^d$ function is hard to distinguish from the $\ln t / t$ if the exponent $d$ is slightly larger than $1$. On the other hand, in Fig.~\ref{fig:kinetics_comparison} (b), we can see that there are straight lines for all but the three most extreme elongations of rectangles, for which we expected a power law kinetics, as due to finite size effects, they correspond to one-dimensional RSA \cite{Swendsen1981, Pomeau1980}. To further support this observation we studied the dependence on the square of Pearson's coefficient of the linear fit for data from panel (b) $r$ on the scaled anisotropy of rectangle $\alpha$. The fitting was done for $ln(t)/t < 0.1$. For almost all anisotropies the dependence is nearly linear as $(1-r^2) \approx 0$. The more significant deviations are observed, as expected, for the largest anisotropies, and interestingly, for scaled anisotropy $\alpha$ corresponding to $f=2$. The latter case could be the result of a quite large range of fitting because the linear dependence of $\theta - \theta(t)$ on $\ln(t)/t$ for parallel squares is reported for $\ln(t)/t < 0.002$ only \cite{Brosilow1991}. In summary, we conclude that in the general case ($\alpha \ll 1$), the RSA kinetics for packings built of rectangles placed on a two-dimensional plane and aligned horizontally and vertically only is governed by the logarithmic law (\ref{eq:log}).
\subsection{Orientational order}
\begin{figure}[ht]
 \centering
 \includegraphics[width=0.8\columnwidth]{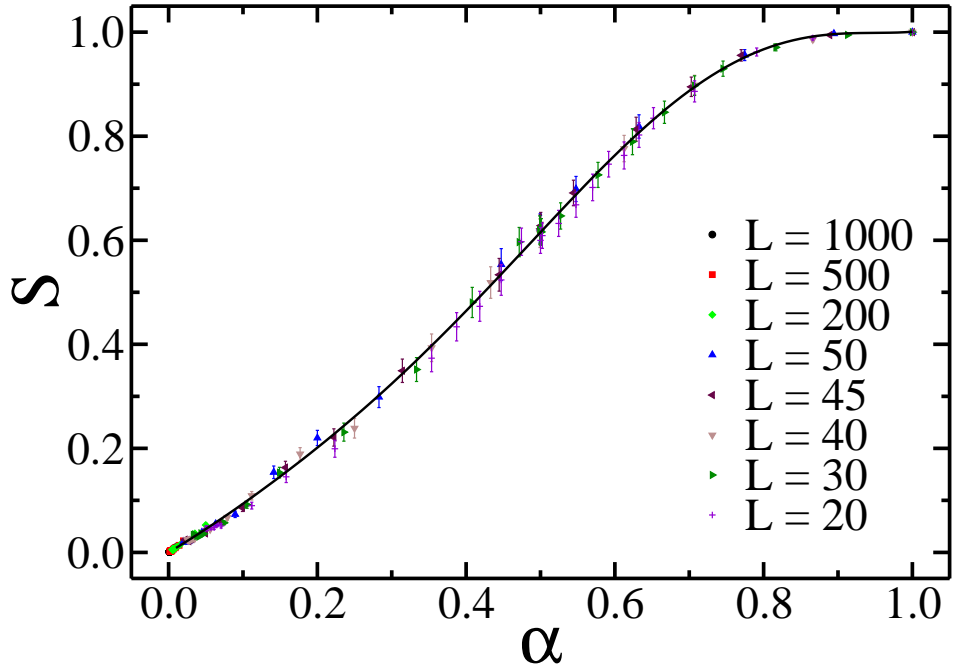}
 \caption{Orientational order parameter $S$ as a function of scaled anisotropy $\alpha$ (Eq.~(\ref{eq:alpha})) for different values of $L$. Different colors and shapes of points correspond to different values of $L$. The value of $S$ for each set of parameters $(\alpha, L)$ was averaged over $100$ independent packings. The solid line is to guide the eye.}
 \label{fig:S_f}
\end{figure}

Orientations in RSA packings are random, thus typically there is no global orientational order. On the other hand, locally anisotropic figures tend to align in parallel due to the smaller excluded surface of such configuration \cite{Sherwood1997, Ciesla2013, Abritta2022}. However, for relatively large values of the scaled anisotropy $\alpha$ (Eq.~(\ref{eq:alpha})), when the values $l$ and $L$ are comparable, all the rectangles have to be aligned in the same direction because perpendicular alignment is impossible due to crossing. For slightly smaller anisotropies if the first few rectangles happen to be in the same orientation then they cause hindrance to the deposition of rectangles in the orthogonal direction, all the following deposited rectangles are forced to be in the same direction. To study this phenomenon, we define an order parameter, as:
\begin{equation}
\label{eq:S}
S = \left| \frac{n_h - n_v}{ n_h + n_v} \right|,
\end{equation}
where $n_h$ and $n_v$ are, respectively, the number of rectangles in the horizontal and vertical direction. So, $S$ is equal to $0$ for a packing where the number of the two orientations of the rectangles is the same and $1$ if only one of them is present (fully ordered packing). It appears that $S$ depends mostly on the value of scaled anisotropy $\alpha$ (Eq.~(\ref{eq:alpha})) and does not depend on particular values of the packing sizes $L$ and rectangles sides-ratio $f$ -- see Fig.~\ref{fig:S_f}. 
As expected, for small $\alpha$, there is no significant preference of one alignment to the other, but with the growth of the relative anisotropy, packings become more orientationally ordered. For $\alpha \to 1$ almost all rectangles in a packing share the same orientation.
\subsection{Percolating clusters}
\begin{figure}[ht]
 \centering
 \includegraphics[width=0.4\columnwidth]{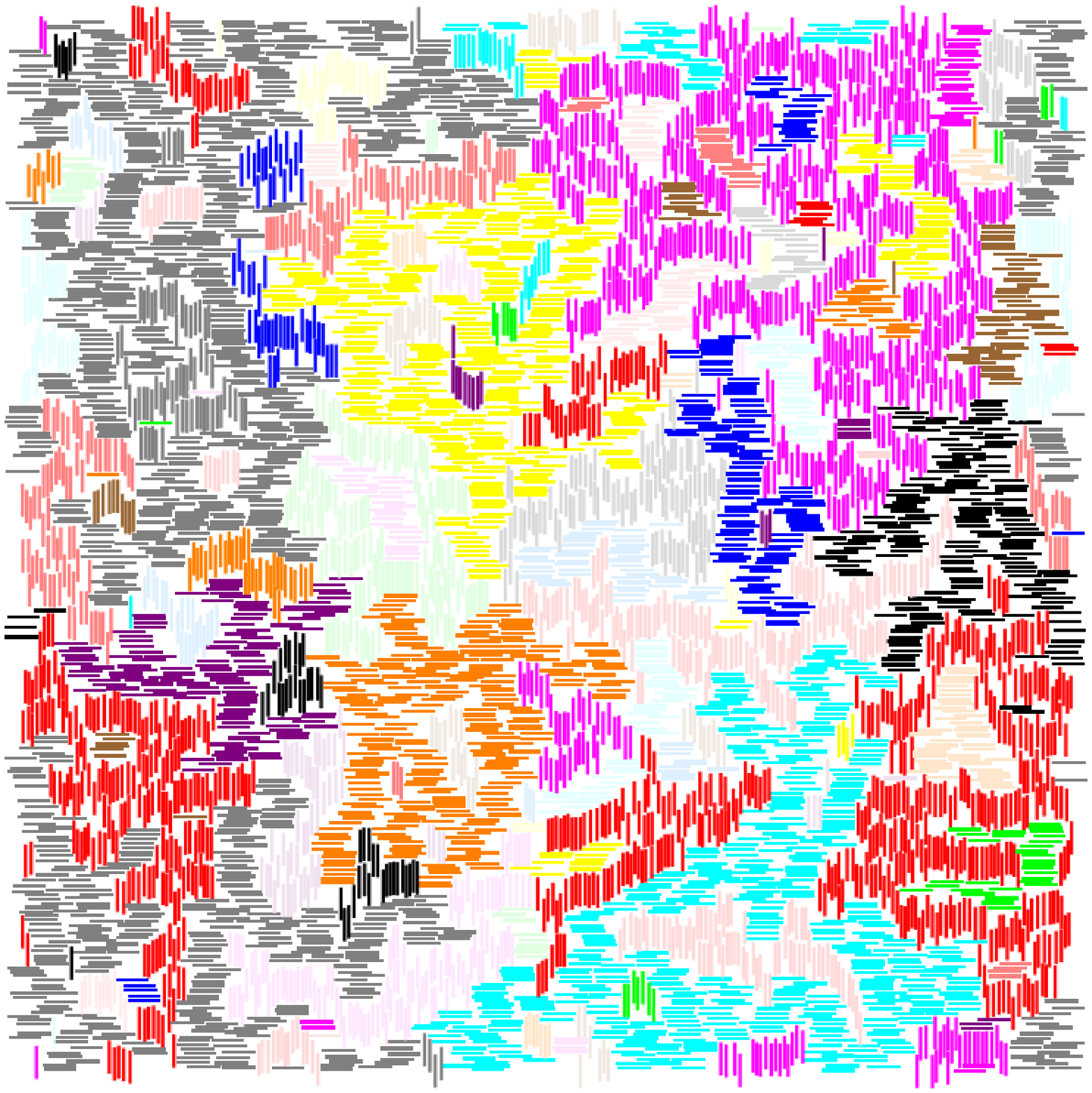}
 \includegraphics[width=0.4\columnwidth]{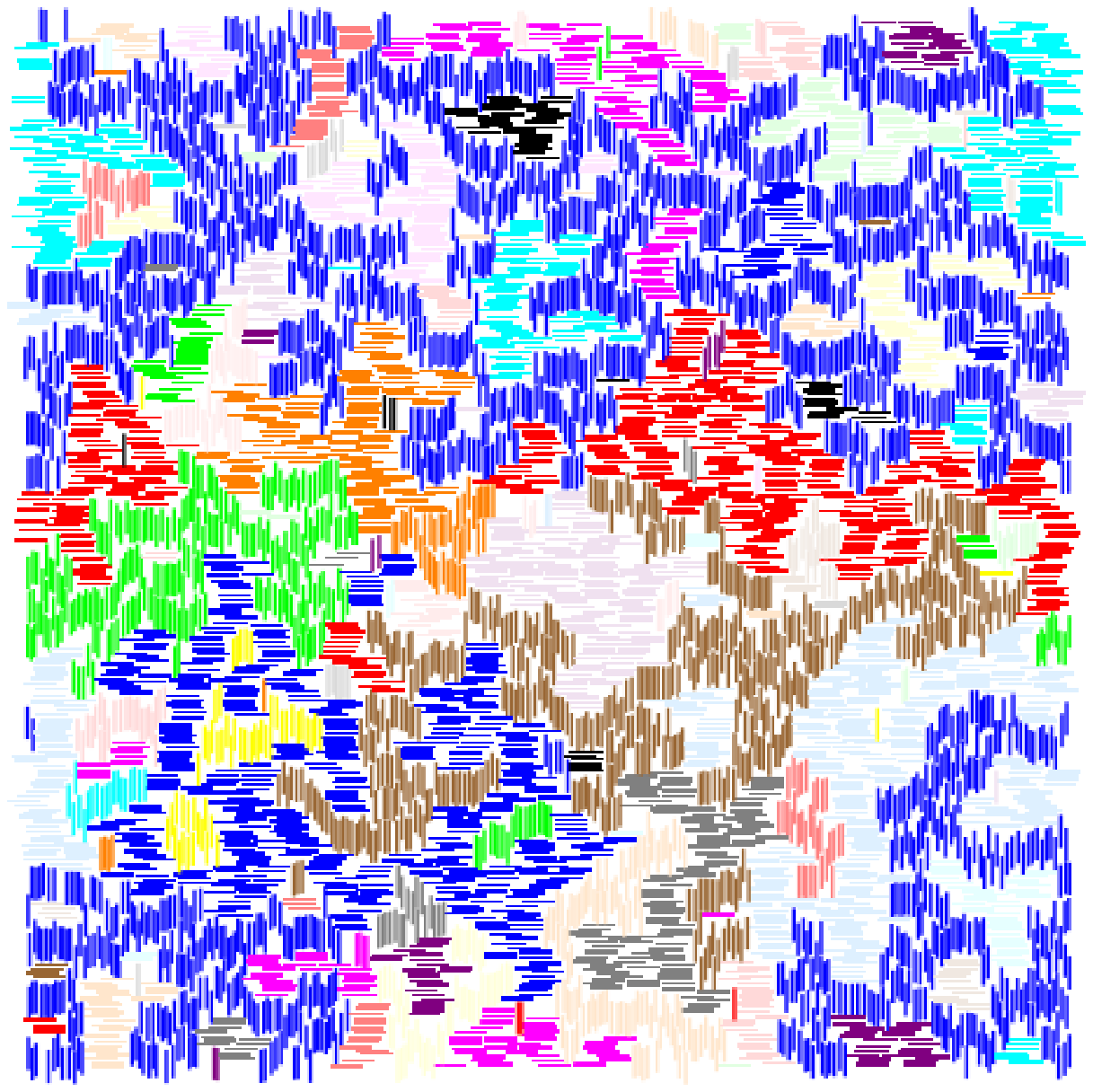}
 \caption{Saturated packing divided into domains. Rectangles in a cluster have the same color. On the left, there is no percolating domain while on the right, the percolating cluster is the blue one. These example packings were generated for $L=100$ and $f=10$ ($\alpha \approx 0.013$).}
 \label{fig:clusters}
\end{figure}

\begin{figure}[ht]
 \centering
 \includegraphics[width=0.8\columnwidth]{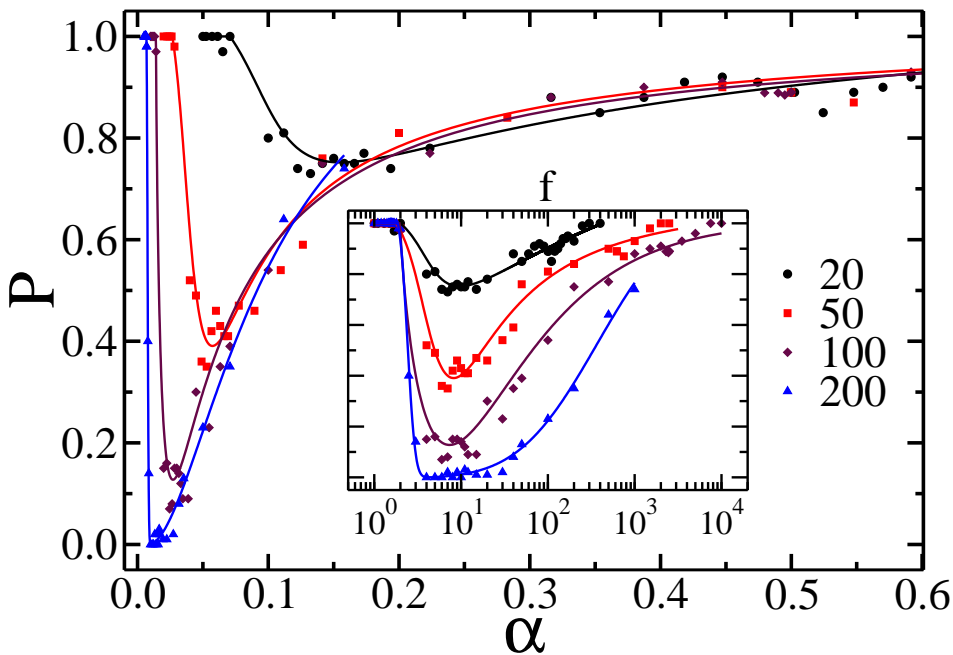}
 \caption{Percolation probability dependence on the scaled anisotropy $\alpha$ (Eq.~(\ref{eq:alpha})) for various packing sizes. Inset shows the same dependence but only on anisotropy $f$ and using a different horizontal scale. Solid lines are to guide the eye. For each set of parameters $f$ and $L$ the probability was estimated by studying $100$ independently generated saturated packings, thus the relative statistical error is of the order of $10\%$.}
 \label{fig:percolations}
\end{figure}

We defined a cluster or a domain as a set of identically oriented rectangles that are so close that no other figures can be placed between them, which means that the edge-to-edge distance measured along the horizontal or vertical axis is not larger than the shorter side of a rectangle. This definition is in analogy with one for $k$-mers in a square lattice \cite{Lebovka2011, Tarasevich2018}, and can be extended to the case of non-saturated packings on a continuous plane. Then the edge-to-edge distance between neighboring shapes has to be smaller than a given arbitrary value. Thus, the domain sizes depend on it \cite{Lebovka2021}.
The example saturated packings divided into such clusters are shown in Fig.~\ref{fig:clusters}.
Here, we are focused on the probability of percolation, i.e., the existence of a cluster that splits the packing into two separated parts -- see Fig.~\ref{fig:clusters}. The results are shown in Fig.~\ref{fig:percolations}.
Interestingly, percolations are highly probable for both: very small and very large rectangles' anisotropy. For squares, because packing is saturated, all the particles are in one domain, thus the percolation always is observed. It does not change for small anisotropy ($f$ only slightly higher than $1$), but then a majority of the figures belong to one of two large domains consisting of horizontally or vertically aligned shapes. At relatively small values of rectangle aspect ratio $f$ ($<10$) the percolation probability drops down with the growth of $f$. The minimums in $P(f)$ dependencies are observed for all values of $L$ at approximately the same value of $f$ ($\approx 10$). The depth of this minimum increases with $L$, which reflects the manifestation of the finite size effects in percolation. Particularly, for large enough values of $L$ ($L=200$) the minimum probability drops down practically to $0$ (Fig.~\ref{fig:percolations}). It is worth noting, that the orientational order propagation in RSA packings built of anisotropic, unoriented figures is typically limited to very short distances -- see e.g. \cite{Kasperek2018}. Thus, in general, the sizes or orientationally oriented domains are small and will not form a percolating cluster, which explains the minimum of the percolation probability observed for small $\alpha$. On the other hand, in the limit of $\alpha\to 1$, when the length of the rectangle $l$ becomes comparable to the system size $L$, the packing becomes ordered (see Fig.~\ref{fig:S_f}), so the majority of shapes are parallel and the probability that they will form one big cluster raises and therefore percolating probability $P \to 1$.
\section{Conclusions}
Finite-size scaling behavior of the density $\theta$ of saturated RSA packings built of horizontally or vertically aligned rectangles is mainly determined by the value scaled anisotropy $\alpha$ defined by Eq. (\ref{eq:alpha}). In the limit of $\alpha \to 0$ (at a relatively small value of aspect ratio $f$ of rectangles and very large size of packings $L$) the finite size scaling is practically absent. The maximum value of $\theta=0.60252694 \pm 0.000019$ at $f\approx 1.56$. The more complicated behavior of $\theta(\alpha)$ is observed at the finite values of $\alpha$ (Fig. \ref{fig:theta_f_long}). It evidently reflects the manifestation of finite-size scaling effects. Particularly, in the limit $\alpha \to 1$ (i.e. at $l\to L$) the problem reduces to RSA in one dimension and the packing fraction rises up to 0.7476... \cite{Renyi1958}. The kinetics of packing growth fits well with the extended logarithmic law (Eq. (\ref{eq:log})) predicted \cite{Swendsen1981} and observed \cite{Brosilow1991} in the case of parallelly aligned squares.  Because of only two possible orientations of rectangles, the packing can exhibit global orientational order that grows with scaled anisotropy $\alpha$. The analysis of rectangles connectivity in saturated packing is performed via the calculation of percolation probability $P$ at different values of $\alpha$. The data reveals the large values of $P \approx 1$ in the limits of $\alpha \to 0$ and $\alpha \to 1$. The minimum of $P$ at some intermediate length-to-width ratio ($f \approx 10$) is detected. 

In general, it is worth noticing that in some aspects the behavior of the system is fully determined by $f$, e.g., the position of the first maximum of packing fraction or the minimum of percolation probability, while other characteristics are governed by $\alpha=\sqrt{f}/L$, like for example the transition saturation packing fraction or the global orientational order. This reflects the manifestation of percolation finite-size scaling effects.  
\section*{Acknowledgments}
Part of the numerical simulations was carried out with the support of the Interdisciplinary Center for Mathematical and Computational Modeling (ICM) at the University of Warsaw under grant no. GB76-1. N.L. acknowledges partial funding from the NAS of Ukraine, Project No. 0123U101080.  
\section*{References}
%
%
%
\bibliographystyle{iopart-num}
\bibliography{main}
\end{document}